\documentclass[10pt,conference]{IEEEtran}
\usepackage{enumitem}
\usepackage{calc}
\usepackage{multicol} 

\usepackage{soul}
\usepackage{microtype}
\usepackage{tabularx}
\usepackage[noend]{algpseudocode}
\usepackage{booktabs}
\usepackage{multirow}
\usepackage{graphicx}
\usepackage{textcomp}
\usepackage{tikz}
\usepackage{xcolor}
\usepackage{enumitem}
\usepackage{listings}
\usepackage{caption}
\usepackage[linesnumbered,ruled,lined]{algorithm2e}
\usepackage{alltt}
\usepackage{multirow}
\usepackage{listings}
\usepackage{color, soul}
\usepackage{textcomp}
\usepackage{framed}
\usepackage{hhline}
\usepackage{subcaption}
\usepackage[breakable]{tcolorbox}
\usepackage{float}
\usepackage{array}
\usepackage{flushend}
\usepackage{makecell}
\usepackage{balance}
\usepackage{centernot}
\usepackage[T1]{fontenc}
\usepackage[flushleft]{threeparttable}
\usepackage{url} 
\usepackage{amsthm}
\newtheorem{definition}{Definition}
\usepackage{stmaryrd}
\usepackage[colorinlistoftodos,textwidth=15mm,textsize=tiny,obeyFinal]{todonotes}
\usepackage{lstautogobble}
\usepackage{siunitx}
\definecolor{gray}{RGB}{215,215,215}
\definecolor{light-gray}{gray}{0.8}
\definecolor{codegreen}{rgb}{0,0.6,0}
\definecolor{codegray}{rgb}{0.5,0.5,0.5}
\definecolor{mygray}{rgb}{0.7,0.7,0.7}
\definecolor{codepurple}{rgb}{0.58,0,0.82}
\definecolor{backcolour}{rgb}{0.95,0.95,0.92}
\def\BibTeX{{\rm B\kern-.05em{\sc i\kern-.025em b}\kern-.08em
    T\kern-.1667em\lower.7ex\hbox{E}\kern-.125emX}}

\newcommand{\xMapsto}[2][]{\ext@arrow 0599{\Mapstofill@}{#1}{#2}}
\def\Mapstofill@{\arrowfill@{\Mapstochar\Relbar}\Relbar\Rightarrow}

\usepackage{fancyvrb}
\usepackage{tcolorbox}
\usepackage{hyperref}
\newcommand{\watertool}{\textsc{Water}}
\newcommand{\edtool}{\textsc{Edit Dis}}

\newcommand{\vistatool}{\textsc{Vista}}
\newcommand{\webevotool}{\textsc{WebEvo}\xspace}
\newcommand{\sftmtool}{\textsc{SFTM}\xspace}

\newcommand{\uitestfix}{\textsc{UITestFix}\xspace}

\let\othelstnumber=\thelstnumber
\def\createlinenumber#1#2{
    \edef\thelstnumber{%
        \unexpanded{%
            \ifnum#1=\value{lstnumber}\relax
              #2%
            \else}%
        \expandafter\unexpanded\expandafter{\thelstnumber\othelstnumber\fi}%
    }
    \ifx\othelstnumber=\relax\else
      \let\othelstnumber\relax
    \fi
}

\newtcbox{\mybox}[1][breakable]{on line, enlarge top by=10pt, enlarge bottom by=10pt,
     boxsep=8pt, boxrule=2pt, size=small, arc=1mm}

\definecolor{main-color}{rgb}{0.6627, 0.7176, 0.7764}
\definecolor{string-color}{rgb}{0.3333, 0.5254, 0.345}
\definecolor{key-color}{rgb}{0.8, 0.47, 0.196}
\lstdefinestyle{mystyle}
{
    language = Java,
    basicstyle = {\ttfamily \color{main-color}},
    stringstyle = {\color{string-color}},
    keywordstyle = {\color{key-color}},
    keywordstyle = [2]{\color{lime}},
    keywordstyle = [3]{\color{yellow}},
    keywordstyle = [4]{\color{teal}},
    morekeywords = [3]{<<, >>},
    morekeywords = [4]{++},
    basicstyle=\ttfamily\scriptsize,
    commentstyle=\color{blue}\ttfamily,
    morecomment=[f][\lstbg{red!20}]-,
    morecomment=[f][\lstbg{green!20}]+,
    morecomment=[f][\lstbg{yellow!20}]++,
    morecomment=[f][\lstbg{yellow!20}]--,
    morecomment=[f][\textit]{@@},
    texcl=false
}

\definecolor{grey}{rgb}{0.7,0.7,0.7}

\newcommand{\lstbg}[3][0pt]{{\fboxsep#1\colorbox{#2}{\strut #3}}}
\lstdefinelanguage{diff}{
  basicstyle=\ttfamily\scriptsize,,
  morecomment=[f][\lstbg{red!20}]-,
  morecomment=[f][\lstbg{green!20}]+,
  morecomment=[f][\lstbg{yellow!20}]++,
  morecomment=[f][\textit]{@@},
  texcl=false
}

\newcommand{\ignore}[1]{}

\title{Understanding and Enhancing Attribute Prioritization in Fixing Web UI Tests with LLMs}

\author{\IEEEauthorblockN{Zhuolin Xu}
\IEEEauthorblockA{
\textit{Concordia University,}
Canada \\
zhuolin.xu@mail.concordia.ca}
\and
\IEEEauthorblockN{Qiushi Li}
\IEEEauthorblockA{
\textit{Concordia University,}
Canada \\
qiushi.li@mail.concordia.ca}
\and
\IEEEauthorblockN{Shin Hwei Tan$^\dagger$}
\IEEEauthorblockA{
\textit{Concordia University,}
Canada \\
shinhwei.tan@concordia.ca}
}

\begin{document}
\maketitle
\begin{abstract}

The rapid evolution of Web UI incurs time and effort in UI test maintenance. Prior techniques in Web UI test repair focus on locating the target elements on the new Webpage that match the old ones so that the corresponding broken statements can be repaired. These techniques usually rely on prioritizing certain attributes (e.g., XPath) during matching where the similarity of certain attributes is ranked before other attributes, indicating that there may be bias towards certain attributes during matching.
To mitigate the bias, we present the first study that investigates the feasibility of using prior Web UI repair techniques for initial matching and then using ChatGPT to perform subsequent matching. Our key insight is that given a list of elements matched by prior techniques, ChatGPT can leverage language understanding to perform subsequent matching and use its code generation model for fixing the broken statements.
To mitigate hallucination in ChatGPT, we design an explanation validator that checks if the provided explanation for the matching results is consistent, and provides hints to ChatGPT via a self-correction prompt to further improve its results. Our evaluation on a widely used dataset shows that the ChatGPT-enhanced techniques improve the effectiveness of existing Web test repair techniques. Our study also shares several important insights in improving future Web UI test repair techniques.

\end{abstract}

\begin{IEEEkeywords}
Web UI Test Repair, Test Maintenance, UI Element Matching
\end{IEEEkeywords}
\pagenumbering{arabic}
\





\section{Introduction}

When developers change the user interfaces (UI) of a Web application due to rapidly changing requirements, the corresponding Web UI tests need to be manually updated for test maintenance.  
To reduce manual efforts in repairing broken Web UI tests, several automated approaches have been proposed
 in academia and industry to automatically fix broken Web UI test cases caused by software evolution
~\cite{water,vista,UITESTFIX}. The key step in the automated repair of Web UI tests is to modify the broken statements containing outdated element locators by matching the element $e_{old}$ in the old version of a Web application with the element $e_{new}$ in the new version ~\cite{whydowe}. 
Prior Web UI test repair techniques mostly rely on a set of Document Object Model (DOM) attributes (e.g., identifiers and XPath)~\cite{water} or visual information~\cite{vista} 
to determine whether the two elements $e_{old}$ and $e_{new}$ match. 
These techniques extract and compute the similarity of this information to select the most similar element as the result of the matching.


Due to numerous attributes available for matching Web UI elements, these techniques may prioritize certain attributes. For example, \watertool{} \cite{water}, a classical Web UI test repair technique, applies a multi-step matching process using different attributes. First, it searches for identical elements by matching five attributes (\emph{id}, \emph{XPath}, \emph{class}, \emph{linkText}, \emph{name}). If unsuccessful, it then finds similar DOM nodes using additional attributes. Specifically, it identifies elements with the same \emph{tagname}, 
and computes their similarity scores 
between $e_{old}$ and $e_{new}$ 
 with the weighted sum of \emph{XPath} and other attributes where it prioritizes \emph{XPath} similarity based on the heuristic that \emph{XPath} of nodes ``should be very similar across versions''~\cite{water}. As the prioritization and the predefined order on matching these attributes are usually based on heuristic of the tool developers, \emph{the matching algorithm may not accurately reflect the evolution of the Web element}, causing inaccuracy during matching, and subsequently failing to repair the broken statement. Hence, it is important to understand the attribute prioritization used by prior test repair approaches. 
 Meanwhile, prior learning-based techniques show promising results in combining different types of information for repairing broken Android GUI tests (e.g., combining word and layout embeddings~\cite{Vifer}, or fusing GUI structure and visual information~\cite{xu2021guider}). The richer representation used by these learning-based techniques has been shown to enhance the accuracy of the UI matching step.     

To solve the aforementioned 
problems of Web UI test repair and to hinge on a richer representation in learning-based approach, 
we present the \emph{first study} of understanding the attribute prioritization and enhancing traditional Web UI test repair approaches with LLMs like ChatGPT. 
Our use of ChatGPT is motivated by the promising results shown in prior studies for solving related software maintenance tasks, (e.g., test generation~\cite{Feldt2023aa}, and automated program repair~\cite{sobania2023analysis,fan2023automated}). 
However, the Web UI test repair problem differs from these tasks as it mainly involves Web element matching where accurate matching results typically lead to the correct repairs. Specifically, our study evaluates the effectiveness of integrating ChatGPT into two representative Web test repair techniques (\watertool{} and \vistatool{} \cite{vista}). To further evaluate the heuristic used in \watertool{} that prioritizes XPath similarity, we also design a simplified variant of \watertool{} that performs matching using only Levenshtein edit distance between XPaths of the old element $e_{old}$ and the new element $e_{new}$ (we call this approach \emph{\edtool}). Our study focuses on Java Selenium tests which are commonly used by Java Web applications. 


The key insights of our approach are twofold: (1) we first rely on traditional Web UI test repair approaches to obtain an initial list of candidate matched elements (which may be biased by the prioritization used by the approach), and then use ChatGPT to perform subsequent matching to further select the best matched element in the candidate list, and (2) as ChatGPT may suffer from the hallucination problem~\cite{Feldt2023aa}, we design our prompt based on OpenAI's official documentation by asking ChatGPT to generate an explanation along with each selection. To mitigate hallucination, we propose an \emph{explanation validator} that automatically checks for the consistency of the explanation, and instruct ChatGPT to self-correct the initial selection based on detected inconsistent explanation.

Our study 
aims to 
answer the following questions: 
\begin{description}[leftmargin=*]
\item[RQ1:] Can ChatGPT help improve the accuracy of Web element matching of prior Web test repair approaches? 

\item[RQ2:] What is the effectiveness of ChatGPT in repairing broken Selenium statements for Web test repair? 


\item[RQ3:] When ChatGPT explains the result of the element matching, what is the quality of the explanation?

\item[RQ4:] Can our proposed explanation validator guide ChatGPT in self-correction to improve the matching and repair results?


\end{description}

\noindent\textbf{\textit{Contributions.}} Our contributions are summarised as follows:
\begin{description}[leftmargin=*]
    \item[Study:] We present the first study of understanding attribute prioritization and enhancing traditional test repair approaches with ChatGPT for Web UI test repair. Our findings include: (1) prior test repair tools have preferences towards certain attributes (e.g., XPath), 
(2) the combination with ChatGPT helps improve the matching accuracy of all evaluated approaches; 
(3) to our surprise, although \edtool{} performs the worst individually, its combination with ChatGPT outperforms all the evaluated approaches in element matching and repair; (4) the repair effectiveness of all evaluated approaches
is generally similar to the matching ones but there is one case where the repair performance decreases; 
(5) our proposed workflow of checking for explanation consistency and generating self-correct prompt as hint to ChatGPT could further improve the effectiveness for certain combinations (e.g., \watertool{}+ChatGPT).
    \item[Technique:] We proposed a novel Web UI test repair technique that uses traditional Web test repair approaches for initial matching and then uses ChatGPT for \emph{subsequent matching} (uses all attributes of each UI element to perform another matching) to further improve the matching accuracy. Instead of prioritizing certain attributes in prior approaches (Table \ref{tab:attribute}), the subsequent matching steps use all attributes. To combat the hallucination problem in ChatGPT, we also design an explanation validator that automatically checks for the consistency of the generated explanation to guide 
    self-correction.  
    \item[Evaluation:] We evaluate three baseline approaches (\watertool{}, \vistatool{}, \edtool{}), their combinations with ChatGPT, and two recent approaches (\webevotool and \sftmtool). Our evaluation on a widely used dataset~\cite{dataset} shows that the proposed workflow further improves the effectiveness of prior approaches in Web UI element matching and test repair.
    
\end{description}

\section{Background And Related Work}
In this section, we introduce related work on test repair to provide a background on existing approaches.

\noindent\textbf{Automatic UI test repair.}
Unlike traditional unit tests, UI tests are typically used to validate the functionality of particular UI application components through simulated user interactions such as button clicks.~\cite{water}.
When a UI application evolves, the corresponding UI tests may crash, leading to significant effort required to manually fix these broken tests.  
Several techniques have been proposed for UI test maintenance~\cite{
water,vista,UITESTFIX,ATOM,CHATEM,sitar,theory_repair,Vifer,color,WTSR}.
Most of these techniques focus on maintaining UI tests for mobile applications~\cite{ATOM,CHATEM,xu2021guider,METER,Vifer}, Web applications~\cite{water,vista,color,UITESTFIX,WTSR} or desktop applications~\cite{sitar,theory_repair}.
Prior Web UI test repair techniques~\cite{water,vista,UITESTFIX} typically (1) execute the test, (2) extract information from the Webpage based on various attributes~\cite{water} or visual information~\cite{vista} of Web elements, (3) use various matching algorithms, and (4) update the locator of the matched element in the UI test to fix the broken test.

Model-based approaches~\cite{ATOM,CHATEM,sitar,theory_repair,dataset} build a model of the application under test and modify the event flow to fix tests.
Meanwhile, several heuristic-based approaches~\cite{water,vista,UITESTFIX} relocate elements with UI matching algorithms and update broken tests via replacement of the matched elements.
As heuristic-based approaches without building a model are more practical for large-scale applications, our paper mainly evaluates these approaches.
A recent approach~\cite{Vifer} first matches elements of two versions of Android apps using various similarity metrics (e.g., semantic embedding
similarity, and layout similarity
based on node embeddings of the GUI layout tree), and then repairs the tests by updating broken locators.

\noindent\textbf{UI element matching.} Given an element from an old version web, UI element matching aims to find the corresponding element in the new version.
UI element matching has been applied to several domains (e.g., test reuse~\cite{10.1145/3460319.3464827,khalili2024semantic}, automated compatibility testing~\cite{CdDiff}, and automated maintenance of UI tests).
While several techniques exist for UI element matching, only a few are suitable for our evaluation.
This paper evaluates classic UI test repair tools \watertool{}~\cite{water} and \vistatool{}~\cite{vista}, which match elements based on attribute information and visual information, respectively. 
Existing repair work mainly relies on strategies based on element information~\cite{water,vista}, or simply combining machine learning algorithms with repair tools~\cite{Vifer}.
COLOR~\cite{color} and GUIDER~\cite{xu2021guider} use a combination of attribute and visual information for matching.
However, COLOR only returns updated locators instead of repairing broken tests while GUIDER focuses on Android test repair. 
Due to the differences between UI tests and existing APR work~\cite{xia2023automated}, we may not be able to directly use prior APR approaches to repair UI tests. 
Unlike prior approaches, we explore the feasibility of using prior test repair approaches to obtain an initial ranked list of elements and use ChatGPT to perform subsequent matching.

Instead of fixing broken Web UI tests, several approaches focus on improving the robustness of the element.
\emph{Robula+} ~\cite{robula2} focuses on generating robust XPath locators because the authors' study found that XPath locators are sometimes the only option. \emph{Robula+} checks whether an element or its ancestors possess unique attributes from a predefined whitelist (e.g., id, name, class). If multiple attributes are found, it chooses the attribute with the highest priority to refine the XPath, thereby enhancing its robustness. 
Expanding on this, subsequent work ~\cite{robula3} recommends multiple locator generators voting on the outcome, favoring reliable generators. 
Our design of the experiment is similar in essence with \emph{Robula+}, letting ChatGPT make a selection among multiple candidates. However, our emphasis is on fixing broken locators instead of improving the robustness of the locators.
While enhancing the robustness of locators is outside the scope of this paper, it could be worthwhile future work to explore LLM-based approaches for generating robust locators. 

\noindent\textbf{LLMs in software maintenance.}
LLMs such as ChatGPT have been applied for various software maintenance tasks ~\cite{msrrepair1,msrgpt,to2023better,Feldt2023aa,sobania2023analysis,fan2023automated}, including related tasks such as (1) test generation~\cite{Feldt2023aa,codamosa,yu2023llm,hu2023augmenting,el2024using,li2023nuances}, Android UI testing~\cite{10172490,feng2023prompting}, predicting flaky tests~\cite{9866550}, and (2) automated program repair~\cite{sobania2023analysis,fan2023automated,msrrepair1,multilingual}. ChatGPT is a transformer-based chatbot developed by OpenAI that helps developers create conversational AI applications~\cite{chatgpt_intro}.
Our proposed approach uses ChatGPT for (1) Web UI element matching, (2) providing explanations for the selected elements, and (3) repairing broken statements in tests.  
To improve code-related generation for LLM-based approaches, a recent approach was proposed using knowledge gained during the pre-training and fine-tuning stage to augment training data and their results show
significant improvement for code summarization and code generation~\cite{to2023better}. Similar to this approach, our proposed workflow also uses the experience gained during the selection of UI elements (i.e., inconsistency in the previous explanation generated by ChatGPT) to improve the UI matching results.
Different from the aforementioned work, this paper focuses on using ChatGPT for solving the Web UI test repair problem that aims to perform UI test update by fixing broken UI element locators. 
To the best of our knowledge, our study also presents the first attempt that investigates and improves the quality of explanations provided by ChatGPT by designing an explanation validator to check the reliability of the explanation to improve the effectiveness of UI matching and test repair. 

\section{Motivating Examples}

We show two examples where the combination of ChatGPT helps to improve the matching and test repair results.

\noindent\textbf{Without self-correction.}
Table \ref{table:motivation_example} shows a target element to be matched by \watertool{} in the old version of the MRBS app, and key candidates from the returned list by \watertool{}. Initially, \watertool{} identifies elements with the same text as the target and returns the first match. However, in this case, two elements share the same value for the ``text'' attribute as the target element, and the second one has more similar XPath, position, and size to the target element than the first match, and 
ChatGPT correctly selected ``Candidate 2'' as the matching result from this ranked list. ChatGPT also explained that it chose this element ``Because they share the most similar attributes: xpath, text, tagName, x, y, width, height.'' (its explanation consistency is 0.8, because all attributes except for XPath are consistent with the selection, and the most similar XPath is ``/html[1]/body[1]'' from another candidate). After matching, our approach fed ChatGPT with the repair prompt, and it successfully repaired the broken statement by updating the locator with XPath from ``Candidate 2''. In this example, with ChatGPT's explanation being 80\% consistent, the subsequent matching step by ChatGPT helps select the correct but lower-ranked candidate, leading to successful repair.





\begin{figure}[!t]
    \centering
    \includegraphics[width=0.9\linewidth]{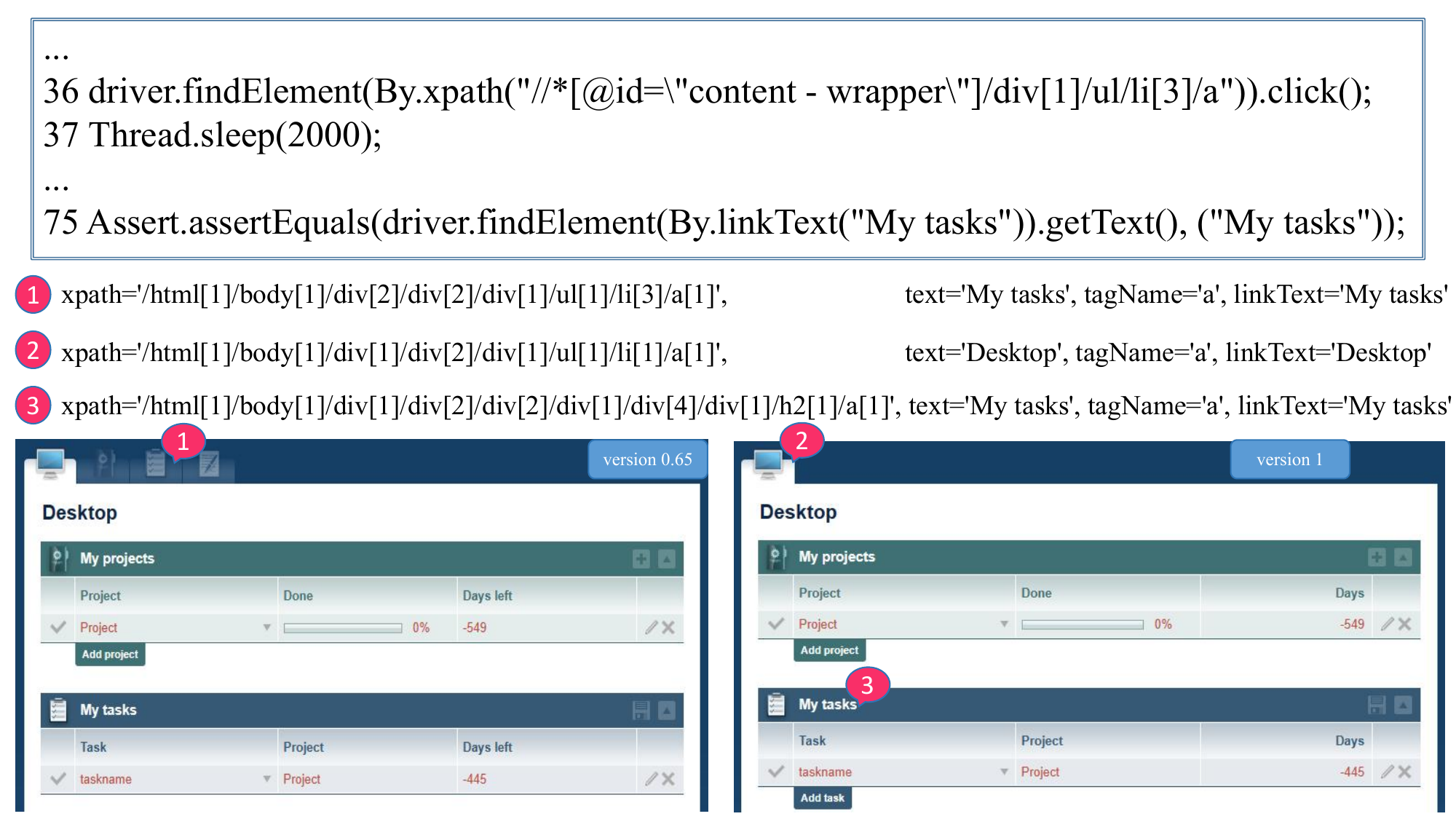}
    \caption{Generated fixes for the broken statement in Collabtive that shows the effectiveness of self-correction.}
    \label{self correct example}
    \vspace{-6pt}
\end{figure}

\noindent\textbf{With self-correction.}  Figure~\ref{self correct example} shows the matched elements for another target element (\textcircled{\raisebox{-0.9pt}{1}}) selected by \watertool+ChatGPT before and after self-correction. Initially, \watertool+ChatGPT selected (\textcircled{\raisebox{-0.9pt}{2}}) with the explanation ``Because they share the most similar attributes: xpath, text, tagName, linkText.''.
However, our explanation validator detects inconsistencies in the attributes \emph{text} and \emph{linkText} (Explanation Consistency=0.5) and provided this feedback via the self-correction prompt. Guided by this, \watertool+ChatGPT revised its selection to (\textcircled{\raisebox{-0.9pt}{3}}) correctly, with a new explanation ``Because they share the most similar attributes: xpath, text, tagName, linkText.'' (Explanation Consistency=0.75 as all attributes except XPath were consistent with the selection). This example highlights how the self-correction mechanism, supported by the explanation validator, guides ChatGPT to identify and fix its inconsistencies to improve matching results.

\begin{table*}
\caption{An example that shows the target element to be matched, and a list of candidate elements returned by \watertool}
\label{table:motivation_example}
\centering
\begin{tabular}{l|l} 
\hline
Target Element & \begin{tabular}[c]{@{}l@{}}\{numericId=96, id='', name='', class='', xpath='/html[1]/body[1]/p[1]', text='Unknown user', tagName='p', linkText='', x=8, y=90,\\ width=1018, height=15, isLeaf=true\}\end{tabular}                                                          \\ 
\hline
Candidate 1    & \begin{tabular}[c]{@{}l@{}}\{numericId=40, id='', name='', class='', xpath='/html[1]/body[1]/div[1]/table[1]/tbody[1]/tr[1]/td[7]/div[1]/a[1]', text='Unknown user',\\ tagName='a', linkText='Unknown user', x=917, y=6, width=111, height=23, isLeaf=true\}\end{tabular}  \\
Candidate 2    & \begin{tabular}[c]{@{}l@{}}\{numericId=47, id='', name='', class='', xpath='\textbf{/html[1]/body[1]/div[2]/p[1]}', text='Unknown user', tagName='p', linkText='', \textbf{x=26, }\\\textbf{y=69, width=982, height=15}, isLeaf=true\}\end{tabular}                        \\
....           & ...                                                                                                                                                                                                                                                                        \\
\hline
\end{tabular}
\end{table*}

\section{Methodology}
\begin{figure}[!t]
    \centering
    \includegraphics[width=0.9\linewidth]
    {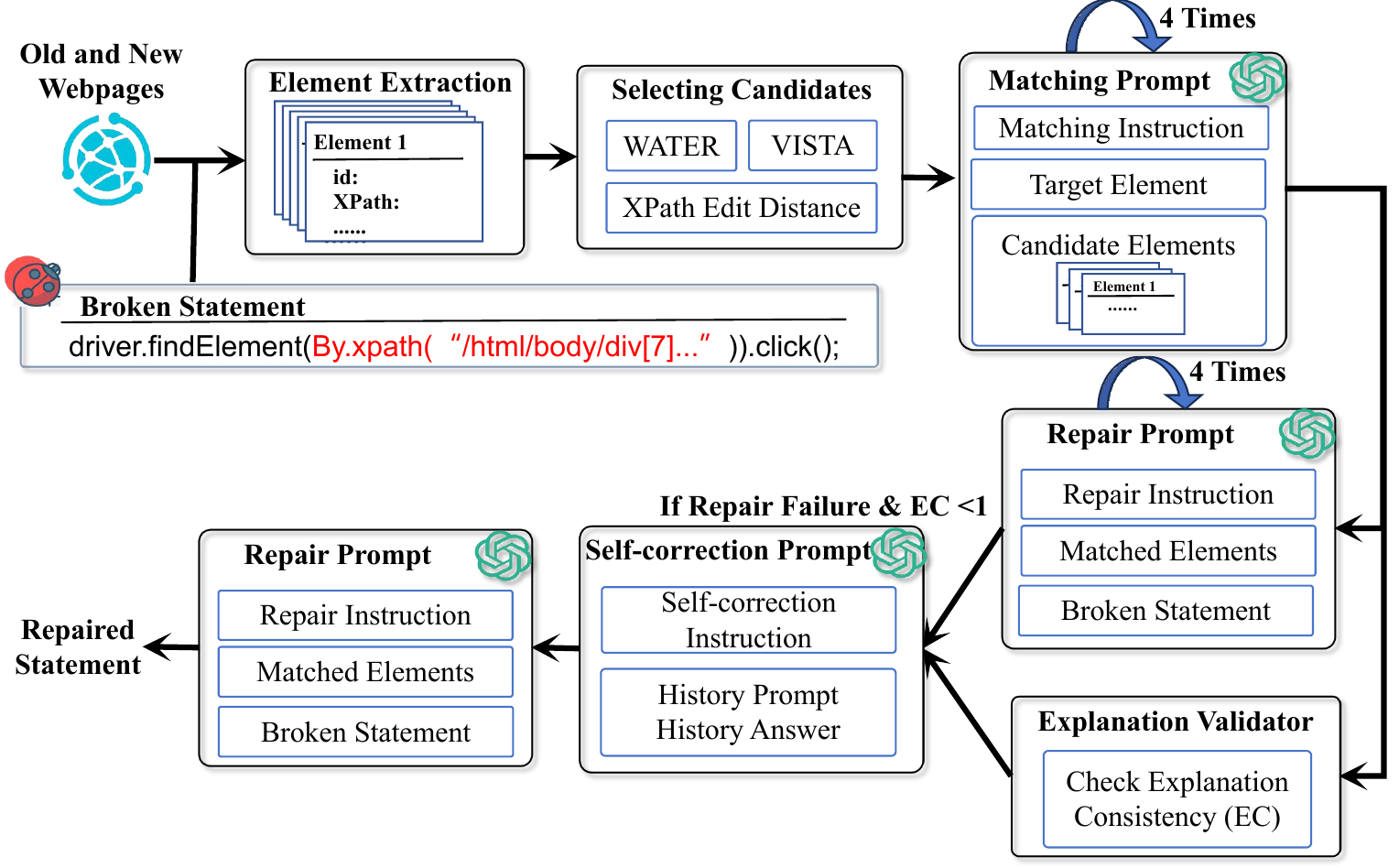}
    \caption{Our proposed workflow of Web UI test repair}
    \label{workflow}
    \vspace{-6pt}
\end{figure}

Figure~\ref{workflow} shows the workflow of our approach. First, our approach extracts information from the target element on the old Webpage based on the broken statement and all elements on the new Webpage. 
Then, it uses a test repair tool (\watertool{}, \vistatool{} or \edtool{}) to rank the elements on the new Webpage, and selects top-ranked ones for ChatGPT to perform further element matching. 
Our approach then instructs ChatGPT to select the matching element and explain its choice by mentioning the most similar attributes it considers. 
Based on the generated explanation, our explanation validator calculates the consistency of the explanation. 
As the accuracy of UI matching cannot be automatically validated without checking if the broken statement has been repaired correctly, our approach generates the repair prompt for ChatGPT to proceed with the repair by instructing ChatGPT to fix the broken statement based on its selected element.
If the repair fails with explanation consistency (EC) less than 1, our approach generates a self-correctness prompt, our approach asks ChatGPT to provide another answer based on the inconsistencies in its original explanation. Finally, we obtain the repaired broken statement from ChatGPT. Due to the randomness of ChatGPT, we rerun the matching and repair process four times, and check if at least one correct repair occurs among the four runs, following the prior study~\cite{sobania2023analysis}. 



\subsection{Prompt Design of Repair and Self-Correction Prompt}
We interact with ChatGPT via the API of the gpt-3.5-turbo model~\cite{gpt3}. 
For optimal results, we design our prompt based on the official OpenAI documentations, including: (1) the official ChatGPT API Guide~\cite{rule_role}
and 
(2) OpenAI six strategies for achieving better results~\cite{rules}. Table~\ref{table:prompt} shows for each sentence of the prompt (the ``Prompt Content'' column), the corresponding rule (the ``Rule in OpenAI official documentation'' column) that inspires the design. The ChatGPT API Guide emphasizes the importance of system instructions in providing high-level guidance for conversations so we use system instructions to design the Web UI test repair context part by telling ChatGPT that (1) it is a web UI test repair tool, and (2) outlining the steps for fixing a broken statement.



Due to token limits, we summarize ChatGPT's matched elements' information from previous dialogues to guide subsequent repair, aligning with the tactic ``For dialogue applications that require very long conversations, summarize or filter previous dialogue.''~\cite{rules}.
Following the rule ``Use delimiters to clearly indicate distinct parts of the input''~\cite{rules}, we use colons to separate labels from respective contents, and curly braces to separate the information of each element, helping ChatGPT distinguish the input. 
Table~\ref{table:prompt} shows our prompt patterns and the applied rules.
The complete prompt generation rules are:
\begin{description}[leftmargin=*]
  \item \textbf{Matching Prompt:} Context{[}p1, p2, p3{]} + Input{[}p8{]}
  \item \textbf{Repair Prompt:} Context{[}p1, p4, p5{]} + Input{[}p9{]}
  \item \textbf{Self-Correction Prompt:} Context{[}p6,  p7{]}
\end{description}


\begin{table*}[h]
\centering
\caption{The patterns of prompt and generation rules}
\label{table:prompt}
\resizebox{0.9\linewidth}{!}{%
\begin{threeparttable}
\centering

\begin{tabular}{lp{2.4cm}p{9.1cm}}
\hline
\multicolumn{1}{l|}{\textbf{PID}} &
  \multicolumn{1}{l|}{\textbf{Rule in OpenAI's official documentation}} &
  \textbf{Prompt Content} \\ \hline
\multicolumn{3}{c}{\textbf{Web UI test repair context patterns: Context}} \\ \hline
\multicolumn{1}{l|}{p1} &
  \multicolumn{1}{l|}{Use system instruction to give high level instructions~\cite{rule_role}} &
  You are a web UI test script repair tool. \\ \hline
\multicolumn{1}{l|}{p2} &
  \multicolumn{1}{l|}{Split complex tasks into simpler subtasks~\cite{split_tasks}} &
  To repair the broken statement, you need to choose the element most similar to the target element from the given candidate element list.\\ 
\multicolumn{1}{l|}{} &
  \multicolumn{1}{l|}{} &
  Give me your selected element's numericId and a brief explanation containing the attributes that are most similar to the target element. \\ \hline
\multicolumn{1}{l|}{p3} &
  \multicolumn{1}{l|}{Provide examples~\cite{rules}} &
  Your answer should follow the format of this example: \\
\multicolumn{1}{l|}{} &
  \multicolumn{1}{l|}{} &
  ‘‘The most similar element's numericId: 1. Because they share the most similar attributes: id, xpath, text.’’ \\ \hline
\multicolumn{1}{l|}{p4} &
  \multicolumn{1}{l|}{Summarize or filter previous dialogue~\cite{rules}} &
  To repair the broken statement, you chose the element \textit{selected element} as the most similar to the target element from the given candidate element list. \\ \hline
\multicolumn{1}{l|}{p5} &
  \multicolumn{1}{l|}{Specify the steps required to complete a task~\cite{rules}} &
  Now based on your selected element, update the locator and outdated assertion of the broken statement. Give the result of repaired statement. \\ \hline
\multicolumn{1}{l|}{p6} &
  \multicolumn{1}{l|}{Use delimiters to clearly indicate distinct parts} &
  This is a previous prompt: \textit{Matching Prompt} \\
\multicolumn{1}{l|}{} &
  \multicolumn{1}{l|}{of the input~\cite{rules}} &
  This is your previous answer: \textit{Corresponding Answer} \\ \hline
\multicolumn{1}{l|}{p7} &
  \multicolumn{1}{l|}{} &
  But your explanation for attributes \textit{attributes} are inconsistent with your selection and this will influence the correctness of your answer. Please answer again. \\ \hline
\multicolumn{3}{c}{\textbf{Input pattern: Input}} \\ \hline
\multicolumn{1}{l|}{p8} &
  \multicolumn{1}{l|}{Use delimiters to clearly indicate distinct parts} &
  Target element: \{\textit{target element}\} \\
\multicolumn{1}{l|}{} &
  \multicolumn{1}{l|}{of the input~\cite{rules}} &
  Candidate elements: \{\textit{candidate element 1}\}, \{\textit{candidate element 2}\} ... \\ \hline
\multicolumn{1}{l|}{p9} &
  \multicolumn{1}{l|}{} &
  Broken statement: \textit{broken statement} \\ \hline
\end{tabular}%
\end{threeparttable}}
\end{table*}

\subsection{Information Extraction and Attribute Prioritization}




\begin{table*}[ht]
\caption{Extracted attributes and the tools that use them. The superscript number indicates the priority of an attribute.}
\centering
\renewcommand{\arraystretch}{1.0}
\vspace{-3pt}
\label{tab:attribute}
\footnotesize
\begin{tabular}{|l|p{9.2cm}|l|}
\hline
Attribute & Description (Example for an element) & Used by Tools \\ \hline
id &  Unique identifier for an element (userName)   & $\watertool^{E1}$ ,$\uitestfix^{E1}$
\\ \hline
name &  Name for an element (submit)  &  $\watertool^{E5}$\\\hline 
 class& Class name for an element (button) &$\watertool^{E3}$
\\\hline 
XPath & Path of an element in the DOM tree 
(/html[1]/body[1]/div[1]/form[1]/
input[1])
&  $\watertool^{E2,L7}$, $\edtool^{L1}$, $\uitestfix^{T2}$\\ \hline
text & Textual information of an element (Enter)  &   $\uitestfix^{L3}$
\\ \hline
tagName & Name of the HTML tag (input)   &  $\watertool^{E6}$, $\uitestfix^{T2}$
\\\hline\hline 
linkText & The text content of a hyperlink (Logout)  &  $\watertool^{E4}$ 
\\ \hline
x & X-coordinate for an element (66) &  $\watertool^8$
\\ \hline
y & Y-coordinate for an element (183)  &  $\watertool^8$
\\ \hline
width & Width of an element (48)  &  $\vistatool^{P1}$, $\watertool^8$
\\ \hline
height & Height of an element (21)  &  $\vistatool^{P1}$, $\watertool^8$
\\ \hline
isLeaf & True if the element is a leaf node of the DOM tree, and false otherwise (true)  &  $\uitestfix^4$ \\ \hline
\end{tabular}

\begin{tablenotes}

\footnotesize
\item{1} The superscript `E' means the tool checks
whether the candidate element’s attribute is exactly the same as
the target element.
\item{2} The superscript `L' indicates the tool uses Levenshtein Distance
to calculate the similarity.
\item{3} The superscript `T' means that the tool uses TF-IDF
to calculate the similarity.
\item{4} The superscript `P' means that the tool calculates the similarity of images, where the width and height of the images affect the calculation. 
\item{5} We refer to the \textbf{VISUAL} mode of \vistatool{}~\cite{vista} for \vistatool{}'s priority, \textbf{DOM} mode of \vistatool{}~\cite{vista} for \watertool{}'s priority. 
\end{tablenotes}
\end{table*}

To extract Web element information, we use an existing component in
 UITESTFIX~\cite{UITESTFIX} to extract relevant information for each element. Specifically, it retrieves the HTML source code of the target Webpage via a Web browser, then uses \emph{Jsoup} library~\cite{jsoup} to extract the Web elements' information in it. 
Note that we did not evaluate UITESTFIX's performance with other tools due to the unavailability of its matching algorithm.
We extract information from several attributes: \emph{id}, \emph{name}, \textit{class}, \emph{XPath}, \emph{text}, \emph{tagName}, \emph{linkText}, \emph{x}, \emph{y}, \emph{width}, \emph{height}, and \emph{isLeaf}.  

\noindent\textbf{Understanding attribute prioritization.} To gain a better understanding of the attribute prioritization used in prior techniques, we read the original paper and the implementation of each technique (if available).
Table~\ref{tab:attribute} shows the attributes and the priorities used by prior Web UI matching and repair approaches. The first two columns of Table ~\ref{tab:attribute} show the descriptions and examples of the attributes. The column ``Used by Tools'' shows if prior approaches 
 $\watertool$, $\vistatool$, $\edtool$, or $\uitestfix$ 
consider these attributes when matching elements and how these tools use them. 
We observe from Table \ref{tab:attribute} that \emph{XPath} is the most commonly used attribute in all approaches. \vistatool{} uses the fewest DOM attributes because it focuses on the image template matching algorithm. Two attributes (ie, \emph{width} and \emph{height}) are used to represent the size of the element. Meanwhile, \watertool{} and \uitestfix{} exclude these two attributes. \watertool{} uses the greatest number of attributes, including \emph{id}, \emph{XPath}, \emph{class}, \emph{linkText}, \emph{name}, \emph{tagName}, \emph{x} and \emph{y}. \uitestfix also uses four other attributes, of which \emph{text} and \emph{isLeaf} are not used by \watertool. \emph{isLeaf} is used by \uitestfix toiteratively refining the similarity tthroughugh the DOM structure.

\subsection{Candidate Selection}

To address token limits, we provide ChatGPT with a list of 10 top-ranked candidate elements from the matching results of prior approaches (\watertool{},\vistatool{}, \edtool{}). We only chose 10 elements due to (1) the model's 4096-token limitation, and (2) the lengthy 
information of each element's 12 attributes. 
We briefly introduce the three approaches below:
\begin{description}[leftmargin=*]
\item[\vistatool{}:] \vistatool{} uses Fast Normalized Cross Correlation algorithm~\cite{openCV} to calculate the template matching results, defining the position and size of the matching area by the element's \emph{x}, \emph{y}, \emph{width}, and \emph{height}. It returns a list where elements of a particular screen position are ranked in descending order based on their similarity scores. Instead of returning only the top 1 element in the original version of \vistatool, we modify \vistatool{} to retrieve the top 10 elements 
as candidates for further matching.
\item[\watertool{}:]

As shown in Table~\ref{tab:attribute}, \watertool{} first checks if any candidate elements share the same \emph{id}, \emph{XPath}, \emph{class}, \emph{linkText}, or \emph{name} as the target elements. and returns the first match. If none are found, \watertool{} calculates similarity score for candidates with the same \emph{tagName} as the target element, and returns the first one exceeding the threshold. The score combines the Levenshtein distance of their \emph{XPath}, and the equivalence of their size and position based on \emph{x}, \emph{y}, \emph{width}, and \emph{height} with a small tolerance for variation.
Notably, \watertool{} gives a greater weight (0.9) to the \emph{XPath} similarity because its authors assume that matched elements have similar \emph{XPaths} after the version update. We modify \watertool{} to return a de-duplicated list of 10 unique candidates instead of a single match.


\item[\edtool{}:] 
Inspired by the idea of prioritizing XPath similarity in \watertool{}, we design a simplified matching algorithm to only consider the XPath similarity. Similar to \watertool{}, we use Levenshtein distance~\cite{edit_distance} to measure the difference between \emph{XPaths}. Levenshtein distance 
measures the minimum number of insertions, deletions, and substitutions needed to transform one string into another.  
Greater distance means that the two elements are less similar. This algorithm has also been used in Web testing (e.g., detecting conflicting and outdated data on Webpages~\cite{levenshtein}). 
This variant returns the top 10 elements ranked in descending order based on their XPath similarities.
\end{description}

\subsection{Explanation Validator}
\label{sec:evalidate}
As shown in the ``Prompt Content'' column in Table~\ref{table:prompt},
we instruct ChatGPT to generate an explanation to describe the attributes used for selecting the best matched element. Our intuition is that \emph{if the provided explanation is consistent, then the selection is more likely to be correct} (and repair is more likely to be successfully generated with correct matches). Based on this intuition, we designed an explanation validator to check if ChatGPT's explanation is consistent with the actual selection. Specifically, for each attribute $a$ mentioned in the explanation, our explanation validator calculates the following to determine the most similar element for the consistency calculation $cons(a_i, R)$ where $cons(a_i, R)$=1 if the most similar element is selected:

\begin{description}[leftmargin=*]
  \item[Screen position:] We use Euclidean distance to compute position-related attributes (e.g., x and y coordinates). We select the element with the minimum distance as the most similar.
  \item[Size:] We use the product of width and height, and consider the one with the smallest size difference as the most similar.
  \item[isLeaf:] We check if the \emph{isLeaf} values (true or false) are same. 
  \item[Other attributes:] We use Levenshtein edit distance 
  to measure similarity, with lower values indicating higher similarity.
\end{description}

In the cases where multiple candidate elements and the target element share the same similarity for a particular attribute, we will retain the results of multiple candidate elements and consider the explanation given by ChatGPT to be consistent if either one of these candidate elements has selected. 

\begin{definition}[Explanation Consistency (EC)]
\vspace{-5pt}
\label{def:ec}
Given the \emph{target element} $t$ (the element in the old version of the Webpage to be matched), the selection result $R$ and an explanation $e$ where $e$ mentioned one or more attributes $A=a_1,a_2,...,a_n$, we calculate the \emph{Explanation Consistency (EC)} of $e$ by computing 
 cons($a_i, R$) for each attribute $a_i$ where $cons(a_i, R)$ checks whether each attribute $a_i$ of $R$ is most similar to that of the target element $t$ ($cons(a_i, R)$=1 if the most similar element based on $a\_{i}$ is selected in $R$, and $cons(a_i, R)$=0 otherwise).


\[
\small
EC(e)=\frac{\sum_{i=1}^{n} cons(a_i\in A, R)}{|A|}
\]
\end{definition}

Def. ~\ref{def:ec} presents the definition for Explanation Consistency (EC). 
For each explanation generated by ChatGPT, our explanation validator checks whether the attributes mentioned in the explanation that are similar between the selected element and the target element are consistent with the calculated values of all mentioned attributes. If the calculated values for all mentioned attributes are consistent, then our explanation validator considers the provided explanation as \emph{consistent} across all mentioned attributes ($EC$=1).

\section{Experimental Setup}
\noindent\textbf{Dataset.}
We use an existing dataset~\cite{dataset} to evaluate the effectiveness of Web test repair approaches. We use this dataset because (1) it is the only publicly available dataset for Web UI test repair, and (2) it has been widely used in prior evaluations of Web UI test repair approaches~\cite{UITESTFIX,dataset}. The dataset contains Java Selenium UI tests from five open-source real-world Web applications (and VISTA~\cite{vista} used 3 of them). All of these open-source applications are hosted in Sourceforge (except for MantisBT that is hosted in GitHub~\cite{mantisbt}). We follow the same filtering process of a prior evaluation~\cite{UITESTFIX} to remove duplicated tests and non-broken tests. Subsequently, we obtained 62 test cases containing 139 broken statements as our dataset. Each test script contains 1 to 6 broken statements, with an average of 70 lines of code per script.

\noindent\textbf{Baselines selection.} We evaluated three approaches (\watertool{}, \vistatool{}, \edtool{}) that are widely used in prior studies of Web UI test repair. 
Similar to prior study~\cite{UITESTFIX}, we exclude the model-based tool~\cite{dataset} as the provided code leads to compilation errors due to missing dependencies. We did not compare against several approaches~\cite{UITESTFIX, WTSR} as their matching algorithms are not publicly available so we could not evaluate their effectiveness in both matching and repair.

\noindent\textbf{Comparison with recent approaches.} 
To assess the effectiveness of our approaches, we also evaluate two recent approaches: \webevotool~\cite{shao2021webevo}, which employs DOM tree-based change detection, history-based semantic structure change detection, and semantics-based visual search to match elements; and \sftmtool{}~\cite{sftm2023}, which uses TD-IDF to calculate the initial similarity score on tokenized attributes and iteratively refines the matching based on the DOM structure.
We did not combine these approaches with ChatGPT due to budget constraints.

We used the implementations of \watertool, \vistatool, \webevotool, \sftmtool{} from \uitestfix\cite{UITESTFIX} for evaluation, which includes the re-implemented \watertool{} by \vistatool\cite{vista}, and open-source implementations for the other tools.


\noindent\textbf{Preparing ground truth dataset.} As our evaluation dataset ~\cite{dataset, UITESTFIX} only has tests for the old versions of apps, and the test fixes are unavailable, we need to manually label the ground truths for the matching UI elements for the new versions. 
Specifically, two annotators independently labeled ground truths for each UI element located in the broken statement in the dataset for the three individual baselines (\watertool,\vistatool,\edtool). The annotators are graduate students with over one year of experience in relevant research of Web UI test repair. For 12 cases out of 3*139=417 cases (139 for each of the 3 individual baselines), the annotators had disagreements and met to resolve. 

Table \ref{table:open-source-application-version} shows the old and updated versions of the open-source Web apps in our dataset. The ``$\Delta$V'' column denotes the number of versions between them where a greater number means more significant UI changes, posing more challenges to matching and repair tasks. The ``Test'' column denotes the number of tests for each app, whereas the ``Broken Stmt'' denotes the number of broken statements per app. Our experiment assumes each broken statement is independent so that we can measure the effectiveness of ChatGPT in fixing each broken statement. This assumption is similar to
prior evaluations~\cite{dearrepair,lutellier2020coconut} of learning-based automated repair techniques where the correct fault location is provided. 


All experiments are run on a computer with an Intel Core i5 processor (\SI{1.6}{GHz}) and \SI{12}{GB} RAM. For the experiments related to ChatGPT, we use the API of gpt-3.5-turbo model (which was the latest at the time of our experiments) and set the temperature to 0.8 as used in prior work~\cite{fan2023automated}. 

\begin{table}[]
\centering
\small

\caption{Statistics of open-source Web apps in our dataset}
\label{table:open-source-application-version}
\resizebox{1\linewidth}{!}{%
\begin{tabular}{l|rrr|rr}
\hline
Application & \multicolumn{1}{c}{$\Delta$V} & \multicolumn{1}{c}{Old Version} & \multicolumn{1}{c|}{Updated Version} & \multicolumn{1}{c}{Tests} & \multicolumn{1}{c}{Broken Stmt} \\ \hline
AddressBook & 8                             & 4.0                             & 6.1                                  & 2                         & 2                               \\
Claroline   & 29                            & 1.10.7                          & 1.11.5                               & 27                        & 53                              \\
Collabtive  & 5                             & 0.65                            & 1                                    & 4                         & 11                              \\
MantisBT    & 38                            & 1.1.8                           & 1.2.0                                & 25                        & 66                              \\
MRBS        & 24                            & 1.2.6.1                         & 1.4.9                                & 4                         & 7                               \\ \hline
Avg/Total   & 21                            & -                               & -                                    & 62                        & 139                             \\ \hline
\end{tabular}
}
\end{table}
\section{RQ1: Effectiveness of UI Matching}
We evaluate the effectiveness of UI matching by calculating (1) the number of correct matches, and (2) ranking performance of the baselines. 


\subsubsection{Matching Result and Analysis}
\label{sec:matchresult}

\begin{table*}[]
\caption{The number of correct matching and repairs of different approaches. Bold values indicate the best performance.}
\setlength{\tabcolsep}{2pt}
\label{table:matching_repair_result}
\resizebox{1\linewidth}{!}{
\centering
\begin{tabular}{l|rrrr|rrrrrrrrrrrr}
\hline
\multirow{3}{*}{Applications} & \multicolumn{4}{c|}{Recent Approaches}                                                                            & \multicolumn{12}{c}{Baselines and Combination Approaches (with self-correction)}                                                                                                                                                                                                                                                                                                       \\ \cline{2-17} 
                              & \multicolumn{2}{c|}{WebEvo}                             & \multicolumn{2}{c|}{SFTM}                               & \multicolumn{2}{c|}{VISTA}                                     & \multicolumn{2}{c|}{VISTA+ChatGPT}                              & \multicolumn{2}{c|}{WATER}                                     & \multicolumn{2}{c|}{WATER+ChatGPT}                              & \multicolumn{2}{c|}{EDIT DIS}                           & \multicolumn{2}{c}{EDIT DIS+ChatGPT}                   \\
                              & \multicolumn{1}{c}{Matching} & \multicolumn{1}{c|}{Fix} & \multicolumn{1}{c}{Matching} & \multicolumn{1}{c|}{Fix} & \multicolumn{1}{c}{Matching} & \multicolumn{1}{c|}{Fix}        & \multicolumn{1}{c}{Matching} & \multicolumn{1}{c|}{Fix}         & \multicolumn{1}{c}{Matching} & \multicolumn{1}{c|}{Fix}        & \multicolumn{1}{c}{Matching} & \multicolumn{1}{c|}{Fix}         & \multicolumn{1}{c}{Matching} & \multicolumn{1}{c|}{Fix} & \multicolumn{1}{c}{Matching} & \multicolumn{1}{c}{Fix} \\ \hline
AddressBook                   & 0                            & \multicolumn{1}{r|}{0}   & \textbf{2}                   & \textbf{2}               & 1                            & \multicolumn{1}{r|}{1}          & 1                            & \multicolumn{1}{r|}{1}           & \textbf{2}                   & \multicolumn{1}{r|}{\textbf{2}} & \textbf{2}                   & \multicolumn{1}{r|}{\textbf{2}}  & 0                            & \multicolumn{1}{r|}{0}   & \textbf{2}                   & \textbf{2}              \\
Claroline                     & 14                           & \multicolumn{1}{r|}{14}  & 15                           & 15                       & 48                           & \multicolumn{1}{r|}{48}         & \textbf{51}                  & \multicolumn{1}{r|}{\textbf{51}} & 18                           & \multicolumn{1}{r|}{18}         & 18                           & \multicolumn{1}{r|}{18}          & 1                            & \multicolumn{1}{r|}{1}   & 47                           & 47                      \\
Collabtive                    & 5                            & \multicolumn{1}{r|}{5}   & 5                            & 5                        & 1                            & \multicolumn{1}{r|}{1}          & 1                            & \multicolumn{1}{r|}{1}           & 10                           & \multicolumn{1}{r|}{10}         & \textbf{11}                  & \multicolumn{1}{r|}{\textbf{11}} & 8                            & \multicolumn{1}{r|}{8}   & 8                            & 8                       \\
MantisBT                      & 53                           & \multicolumn{1}{r|}{53}  & 44                           & 44                       & 11                           & \multicolumn{1}{r|}{11}         & 37                           & \multicolumn{1}{r|}{37}          & 50                           & \multicolumn{1}{r|}{50}         & 56                           & \multicolumn{1}{r|}{56}          & 34                           & \multicolumn{1}{r|}{34}  & \textbf{64}                  & \textbf{63}             \\
MRBS                          & 2                            & \multicolumn{1}{r|}{2}   & 5                            & 5                        & \textbf{7}                   & \multicolumn{1}{r|}{\textbf{7}} & \textbf{7}                   & \multicolumn{1}{r|}{\textbf{7}}  & 1                            & \multicolumn{1}{r|}{1}          & 4                            & \multicolumn{1}{r|}{4}           & 0                            & \multicolumn{1}{r|}{0}   & 2                            & 2                       \\ \hline
Total                         & 74                           & \multicolumn{1}{r|}{74}  & 71                           & 71                       & 68                           & \multicolumn{1}{r|}{68}         & 97                           & \multicolumn{1}{r|}{97}          & 81                           & \multicolumn{1}{r|}{81}         & 91                           & \multicolumn{1}{r|}{91}          & 43                           & \multicolumn{1}{r|}{43}  & \textbf{123}                 & \textbf{122}            \\ \hline
\end{tabular}
}
\end{table*}

We evaluate a total of eight matching
approaches, including individual baselines (\vistatool, \watertool, \edtool), their combinations with ChatGPT, and two recent approaches. For each approach, we recorded its abilities to correctly match the ground truth elements of the broken statements in our dataset. 
The ``Matching'' columns in Table~\ref{table:matching_repair_result} show the results on the evaluated applications. 
Comparing the overall matching results of individual baselines with their combination with ChatGPT (with self-correction), we observe that all combinations outperform the individual baseline (i.e., 97 versus 68 for \vistatool, 91 versus 81 for \watertool, and 123 versus 43 for \edtool{}). This result confirms our hypothesis that combining prior test repair approaches with ChatGPT help improve the UI matching results. 
Across projects, the improvement is greatest in MantisBT with \vistatool+ChatGPT and \watertool+ChatGPT, in Claroline with \edtool{}. Meanwhile, \edtool{}+ChatGPT outperformed \edtool{} in most projects, including AddressBook, Caroline, and MRBS. For the recent approaches, Table~\ref{table:matching_repair_result} shows that the \webevotool and \sftmtool{} give similar matching results (74 and 71) that are slightly better than the two individual baselines (\vistatool{} and \edtool{}). However, both recent approaches perform worse than all the combination approaches, indicating that the combination with LLMs like ChatGPT can further enhance the matching accuracy of traditional approaches (\watertool{} and \vistatool{}). 



\begin{tcolorbox}[left=0pt,right=0pt,top=0pt,bottom=0pt]

\textbf{Finding 1:} All combinations of prior Web test repair approaches with ChatGPT performs generally better than the corresponding standalone approach (without ChatGPT). 

\textbf{Implication 1:} Our suggested workflow of using prior repair tools for selecting candidate elements and then ChatGPT for subsequent matching help improve the matching accuracy. 

\end{tcolorbox}
We notice that \vistatool{}'s matching algorithm using visual information is effective for certain apps (e.g., Claroline and MRBS). However, \edtool{}+ChatGPT performs the best among all approaches, fixing 122 out of a total of 139 broken statements. 
Specifically, \edtool{}+ChatGPT yields the greatest improvement (186\%) over the individual \edtool{} approach. After combining with ChatGPT, it correctly matches 80 elements that were originally mismatched and ensures that nearly all initially correctly matched statements remain correct (42).
Given that the individual \edtool{} performs the worst among all baselines,
we think this result is counterintuitive as one would select the tool that performs well individually (i.e., \watertool) to combine with ChatGPT to get more improvement.  


\begin{tcolorbox}[left=0pt,right=0pt,top=0pt,bottom=0pt]
\textbf{Finding 2:} Although the individual \edtool{} approach performs the worst among all individual baselines, its combination with ChatGPT outperforms all evaluated approaches. 

\textbf{Implication 2:} \edtool{} 
combines well with ChatGPT. 
By prioritizing only XPath similarity, it is more effective in guiding ChatGPT for subsequent matching. 
\end{tcolorbox}

\noindent\textbf{Effectiveness of the standalone ChatGPT (ChatGPT only).} 
Another natural baseline approach will be to use the standalone ChatGPT for performing all the matching and repair steps. Hence, we also check the effectiveness of standalone ChatGPT by providing all UI elements as candidates instead of using a selection algorithm to choose 10 of them. However, all the prompts throw errors due to the token limit being exceeded for all cases. This is expected as for a Webpage in our dataset, there are an average of 224 UI elements to be matched where attribute information for each element occupies around 101 tokens. The average number of tokens (224*101=21733) also shows the high cost of the standalone ChatGPT.

\subsubsection{Ranking performance of the baselines}
\label{sec:rankresult}
As the combination of \edtool{}+ChatGPT outperforms all other approaches, we investigate the reasons behind the improvement. Specifically, as \watertool{} and \vistatool{} originally return only one element as the best matching result, we analyze the ranking performance of each baseline approach.
Given a selected element $s_e$ by a tool $t$ and the correct element $t_e$ (i.e., the corresponding element in the ground truth), we consider $t$ \emph{hits} if $s_e$ is exactly the same as $t_e$. 
If one of the elements in ranked list produced by a tool $t$ hits, 
we record its ranking to evaluate the ranking performance.
\begin{figure}[!t]
    \centering
    \includegraphics[width=0.95\linewidth]{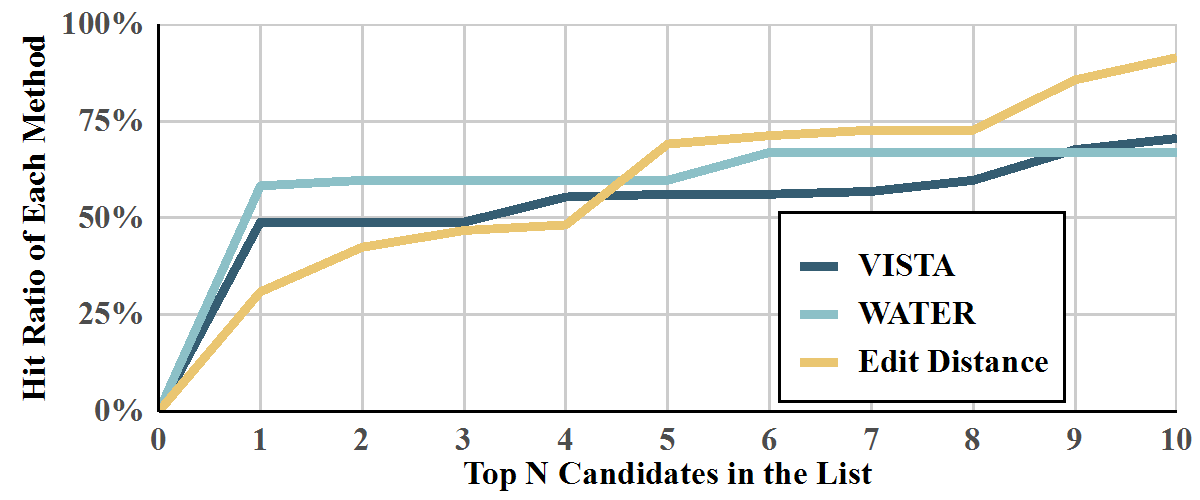}
    
     \caption{Comparison of Hit Ratios for the three baselines}
    \label{Hit Rates}
\end{figure}
We employ metrics commonly used in evaluating top-N recommendation task~\cite{Hit_Ratio_old}: Top-K Hit Ratio (HR, or Recall, in this study, the proportion of experimental instances in which the top N candidates selected by each baseline contain ground truth in the candidate list). Figure ~\ref{Hit Rates} depicts the variation of Hit Ratio with increasing values of N for the three baselines. 
We observe several interesting trends in Figure~\ref{Hit Rates}:  1) When N=1, the Hit Ratio for \vistatool{} and \watertool{} is already around 50\%, while \edtool{}'s is only 30.9\%. This indicates that the top candidate of \vistatool{} and \watertool{} already effectively hits ground truth, whereas \edtool's advantage is less evident. 2) As N increases, the Hit Ratios for \vistatool{} and \watertool{} increase slowly. Even at N=10, their Hit Ratios only increase by less than 25\% compared to when N is 1. In contrast, \edtool{}'s Hit Ratio gradually increases, reaching 68.3\%, 84.2\%, and 90.1\% when N is 5, 9, and 10, respectively. Hence, if expanding the number of selected candidates to the top 10 in the candidate list, \edtool{} is more likely to hit the ground truth element compared to \vistatool{} and \watertool{}, leading to more potential improvement when combined with ChatGPT.

\section{RQ2: Effectiveness and Efficiency of Repair}
Before checking for the repair correctness of each repaired statement, we first check if ChatGPT has the correct matching result for the broken statement because a correct repair can only be generated after a correct matching. 

\noindent\textbf{Repair Correctness.} As the ground truth repaired statements written by developers of the Web apps are unavailable in our dataset, and the repaired statements generated by ChatGPT may have diverse but semantically-equivalent fixes, we need to manually validate the correctness of all generated repaired statements. To reduce the manual effort in validation, we use a semi-automated approach. Specifically, given the original broken statement $o$, the repaired statement $r$, and the ground truth element $e$, we design a parser that automatically parses the locator type (e.g., \texttt{By.name} and \texttt{By.xpath}), and the expression within the locator (e.g., the XPath value) in the repaired statement $r$ to verify their correctness with respect to the ground truth element $e$. For example, if ChatGPT uses \texttt{By.xpath} as the locator, it should use the XPath information of $e$ rather than the value of other attributes. Our parser also identifies cases requiring manual analysis where more substantial changes have been introduced by ChatGPT, including (1) different types of locators between $o$ and $r$, and (2) additional statements added to the repaired statement $r$. 

The ``Fix'' columns in
Table \ref{table:matching_repair_result} shows the repair results for all approaches. The results of the combination approaches (with self-correction) are not only the best results 
across the four runs, on average, \vistatool{}+ChatGPT, \watertool{}+ChatGPT, and \edtool+ChatGPT achieve success in 2.17, 2.01, and 2.89 runs out of 4, respectively, indicating consistent performance across different runs.
From Table \ref{table:matching_repair_result}, we can observe that the number of correct repair is almost equal to the number of correct matching, except for one case in \edtool+ChatGPT. 
For MantisBT, \edtool+ChatGPT correctly matches a element but generates a repair that modifies the original intention of the broken statement. Specifically, the broken statement sends an empty text input ``'' to the element, but \edtool's repaired statement sends ``Test'' to the element. 
Nevertheless, \edtool+ChatGPT still excels with the best performance (122 correct repairs).

\noindent\textbf{Repair Efficiency.} 
To assess potential delays from using ChatGPT, we measured the total time for matching and repair for each broken statement.
 On average, \webevotool takes \SI{45.73}{s}, \sftmtool{} takes \SI{47.98}{s}, \vistatool{} takes \SI{36.91}{s}, \vistatool{}+ChatGPT \SI{41.44}{s}, \watertool{} takes \SI{30.22}{s}, \watertool+ChatGPT \SI{35.11}{s}, \edtool{} \SI{24.02}{s}, and \edtool{}+ChatGPT \SI{28.17}{s} per broken statement. Approaches involving \vistatool{} and \webevotool are more time-consuming because \vistatool{} uses computer vision techniques for matching, which takes more time than matching textual information. \sftmtool{} takes more time because it iteratively propagates and refines the similarity scores during matching process. Overall, ChatGPT adds minimal overhead.

\begin{tcolorbox}[left=0pt,right=0pt,top=0pt,bottom=0pt]

\textbf{Finding 3:} The repair effectiveness is mostly similar to the matching effectiveness (except for one incorrect repair). 

\textbf{Implication 3:} Using correctly matched elements, most approaches could repair correctly. 

\end{tcolorbox}

\section{RQ3: Quality of ChatGPT's Explanation}
\begin{table*}[ht]
\centering
\setlength{\tabcolsep}{1.4pt}
\small
\caption{Effectiveness of SC Mechanism: ``M'' represents the number of times an attribute is mentioned and ``C'' denotes the number of times where the mentioned attribute is consistent in the generated explanation for each approach}
\label{table:explanation}
\resizebox{1\linewidth}{!}{%
\small
\begin{tabular}{l|rr|rr|rr|rr|rr|rr|rr|rr|rr|rr}
\hline
\multirow{2}{*}{Approach} & \multicolumn{2}{c|}{id}                        & \multicolumn{2}{c|}{name}                      & \multicolumn{2}{c|}{class}                     & \multicolumn{2}{c|}{XPath}                     & \multicolumn{2}{c|}{text}                      & \multicolumn{2}{c|}{tagName}                   & \multicolumn{2}{c|}{linkText}                  & \multicolumn{2}{c|}{position}                  & \multicolumn{2}{c|}{size}                      & \multicolumn{2}{c}{isLeaf}                    \\ \cline{2-21} 
                          & \multicolumn{1}{c}{C} & \multicolumn{1}{c|}{M} & \multicolumn{1}{c}{C} & \multicolumn{1}{c|}{M} & \multicolumn{1}{c}{C} & \multicolumn{1}{c|}{M} & \multicolumn{1}{c}{C} & \multicolumn{1}{c|}{M} & \multicolumn{1}{c}{C} & \multicolumn{1}{c|}{M} & \multicolumn{1}{c}{C} & \multicolumn{1}{c|}{M} & \multicolumn{1}{c}{C} & \multicolumn{1}{c|}{M} & \multicolumn{1}{c}{C} & \multicolumn{1}{c|}{M} & \multicolumn{1}{c}{C} & \multicolumn{1}{c|}{M} & \multicolumn{1}{c}{C} & \multicolumn{1}{c}{M} \\ \hline
VISTA+ChatGPT             & 10                    & 35                     & 21                    & 50                     & 112                   & 146                    & 393                   & 590                    & 302                   & 497                    & 368                   & 424                    & 115                   & 135                    & 85                    & 217                    & 131                   & 218                    & 175                   & 188                   \\
WATER+ChatGPT             & 14                    & 25                     & 40                    & 75                     & 116                   & 152                    & 276                   & 547                    & 362                   & 498                    & 360                   & 367                    & 180                   & 203                    & 157                   & 241                    & 226                   & 241                    & 212                   & 215                   \\
EDIT DIS+ChatGPT          & 25                    & 28                     & 24                    & 29                     & 123                   & 144                    & 326                   & 523                    & 362                   & 467                    & 267                   & 309                    & 126                   & 160                    & 137                   & 251                    & 186                   & 249                    & 191                   & 197                   \\ \hline
Total                     & 49                    & 88                     & 85                    & 154                    & 351                   & 442                    & 995                   & \textbf{1660}          & \textbf{1026}         & 1462                   & 995                   & 1100                   & 421                   & 498                    & 379                   & 709                    & 543                   & 708                    & 578                   & 600                   \\ \hline
\end{tabular}
}
\end{table*}



As our study uses explanation to overcome hallucination, it is important to investigate its quality.
To assess the quality of ChatGPT's explanations for element matching results, we use two metrics: (1) \emph{mention frequency} (the number of times where an attribute $a$ has been mentioned $M$), and (2) \emph{mention consistency} (the number of times where an attribute $a$ has been consistently mentioned $C$ where the consistency is determined by our explanation validator described in Section~\ref{sec:evalidate}). These two metrics helps in answering the research questions below:

\begin{description}[leftmargin=*]
\item[RQ3a:] What are the frequently mentioned attributes in ChatGPT's explanation?
\item[RQ3b:] What are the mentioned attributes that are consistent in ChatGPT's explanation?
\end{description}

Table~\ref{table:explanation} presents the mention frequency and mention consistency of ChatGPT for each attribute. Table~\ref{table:explanation} shows that ChatGPT mentions two attributes most frequently: \emph{XPath} (1660) and \emph{text} (1462), whereas the least mentioned attribute is \emph{id} (88). Compared to the priority imposed by prior test repair approaches (shown in Table~\ref{tab:attribute}), this result shows that ChatGPT has certain preferences towards particular attributes since it often mention the \emph{XPath} and \emph{text} attributes regardless of the baseline used for selecting the list of candidate elements.
\vspace{-0.1cm}
\begin{tcolorbox}[left=0pt,right=0pt,top=0pt,bottom=0pt]
\textbf{Finding 4:} ChatGPT frequently mentions the  the \emph{XPath} and \emph{text} attributes in the provided explanations.

\textbf{Implication 4:} Similar to other approaches that have certain priority, ChatGPT 
prioritizes
\emph{XPath} and \emph{text} attributes. 
\end{tcolorbox}

Table~\ref{table:explanation} also shows the attributes with the greatest mention consistency are \emph{text} (1026), \emph{XPath} (995), and \emph{tagName} (995). 
Although ChatGPT prefers using \emph{XPath} (1660) and \emph{text} (1462) for matching, our results show that
prioritizing the \emph{text} attribute over the \emph{XPath} attribute will lead to better matching results (as the \emph{text} attribute is mentioned more consistently).

\begin{tcolorbox}[left=0pt,right=0pt,top=0pt,bottom=0pt]

\textbf{Finding 5:} The most frequently mentioned attribute by ChatGPT (i.e., XPath) tends to lead to incorrect matching (low mentioned consistency). In contrast, the \emph{text} attribute is high in mention frequency and mention consistency.

\textbf{Implication 5:} Prioritizing the \emph{text} attribute over \emph{XPath} attribute is better for ChatGPT as it will be more accurate. 
\end{tcolorbox} 


\noindent\textbf{Correlation between EC and correctness of the matching.} 
Our explanation validator measures the explanation consistency (EC) as a mechanism to check and improve the reliability of ChatGPT's matching results. However, even if our explanation validator can accurately assess EC, it does not guarantee correct matching (i.e., retrieving the target element in the labeled ground truth). To investigate the correlation between our proposed EC and the \emph{correctness of the final matching result} (i.e., whether it selects the ground truth element), we measure the correlation between these two variables. 
As the two variables are binary (correctness) and continuous (EC between 0 and 1) categories respectively, we compute the Point-Biserial Correlation Coefficient ($r_{pbi}$)~\cite{tate1954correlation} for all explanations. As this is a special case of the Pearson Correlation, we assess the strength of the relationship by 
calculating
the correlation coefficient. 

For the three combinations with ChatGPT, the values for the Point-Biserial Correlation Coefficient are: $r_{pbi}$=0.51 for \vistatool{}+ChatGPT, $r_{pbi}$=0.84 for \watertool{}+ChatGPT, and $r_{pbi}$=0.49 for \edtool{}+ChatGPT. These values indicate that EC and the matching correctness for \vistatool{}+ChatGPT and \edtool{}+ChatGPT are only moderately correlated but \emph{strongly correlated for the \watertool{}+ChatGPT combination}. 



 \begin{tcolorbox}[left=0pt,right=0pt,top=0pt,bottom=0pt]

\textbf{Finding 6:} 
EC for \watertool{}+ChatGPT strongly correlates with matching correctness among all combinations.

\textbf{Implication 6:}
The strong correlation shows that our explanation validator 
is the most effective for \watertool{}+ChatGPT to improve its matching correctness.

\end{tcolorbox}
\section{RQ4: Ablation study for the SC mechanism}


\begin{table}[]
\setlength{\tabcolsep}{1.2pt}
\caption{The results before and after self-correction (SC)}
\label{table:scr_result}
\small
 \resizebox{\linewidth}{!}{%
\begin{tabular}{|l|cr|cr|cr|}
\hline
\multirow{2}{*}{Approach} & \multicolumn{2}{c|}{\vistatool{}+ChatGPT} & \multicolumn{2}{c|}{\watertool{}+ChatGPT} & \multicolumn{2}{c|}{\edtool{}+ChatGPT} \\ \cline{2-7} 
                          & before SC                & \multicolumn{1}{c|}{after SC}   & before SC                & \multicolumn{1}{c|}{after SC}   & before SC               & \multicolumn{1}{c|}{after SC} \\ \hline
Matching                  & \multicolumn{1}{r}{97}   & 97                              & \multicolumn{1}{r}{86}   & 91                              & \multicolumn{1}{r}{122} & 123                           \\
Repair                    & \multicolumn{1}{r}{97}   & 97                              & \multicolumn{1}{r}{86}   & 91                              & \multicolumn{1}{r}{121} & 122                           \\ \hline
\end{tabular}
}
\end{table}

As the self-correction (SC) mechanism was the key in the design of our proposed combination, we conduct an ablation study in RQ4 to evaluate the effectiveness of this mechanism. 
Specifically, we count and compare the number of correct matches and repairs before and after the self-correction.
Table \ref{table:scr_result} presents the results of the three approaches combined with ChatGPT before and after self-correction (SC), comparing the correct matches and average explanation consistency (EC). 
Except for \edtool+ChatGPT, the number of correct matches has increased for the other approaches. Notably, \watertool{}+ChatGPT shows the most significant improvement, gaining 5 more correct matches and repairs after self-correction (SC). Meanwhile, we think that \edtool+ChatGPT does not show any improvement due to the moderately positive correlation between EC and matching accuracy (0.49). 

\ignorespaces

\ignorespaces
  \begin{tcolorbox}[left=0pt,right=0pt,top=0pt,bottom=0pt]

\textbf{Finding 7:} The improvement given by self-correction varies across tools. Among the three combinations, \watertool{} benefits the most (5 more correct matching after correction).

\textbf{Implication 7:}
Our proposed workflow of 
guiding ChatGPT via self-correct prompt 
improves the effectiveness of certain combinations (e.g., \watertool{}+ChatGPT). 

\end{tcolorbox}

\section{Implications and Discussions}
Our study identifies several key implications and suggestions for future test repair and ChatGPT research.

\noindent \textbf{Prioritization of attributes by Web UI test repair tools.} Table~\ref{tab:attribute} shows varying attribute prioritization in element matching among prior test repair tools  (e.g., \watertool{} prioritizes XPath similarity, whereas \vistatool{} uses the position and the size information for visual matching). Our study shows that further matching with ChatGPT can improve over prior approaches (Finding 1) by mitigating their bias, leading to more accurate matching. In fact, our study of the frequently mentioned attributes in ChatGPT's explanation also reveals that ChatGPT has preferences towards certain attributes, e.g., \emph{XPath} and \emph{text} (Finding 5). 
Although this paper only studies the prioritization of attributes in two traditional Web test repair approaches, similar biases may exist in other UI matching techniques. In the future, it is worthwhile to study (1) the characteristics of the selected prioritization in other tasks where UI matching algorithms are used (e.g., compatibility testing~\cite{CdDiff}), and (2) improving the effectiveness of other UI matching techniques via subsequent matching using LLMs.  

\noindent\textbf{Test repair techniques used for element pre-selection.} 
Our study that compares three approaches (\watertool{},\vistatool{}, and \edtool{}) for the pre-selection of candidate elements shows that a simplified version of \watertool{} (i.e., \edtool{}) combines well with ChatGPT where \edtool{}+ChatGPT outperforms all the evaluated approaches (Finding 2). Compared to \watertool{} that matches multiple attributes (e.g., \emph{id}, \emph{XPath}) and \vistatool{} that uses visual information for matching, 
\edtool{} that solely relies on \emph{XPath} is less effective as a standalone matching algorithm (it performs worst among all individual baselines). However, our study shows that by using only XPath similarity, \edtool{} delegates the responsibility of matching using other attributes to ChatGPT which later performs subsequent matching. Intuitively, one may think that \watertool{} that performs the best among the individual baselines would lead to the best results when combining with ChatGPT. \emph{As the best individual baseline may not be the most effective combination with ChatGPT}, our study urges researchers to perform thorough evaluation to choose appropriate baseline to combine with ChatGPT for solving other software maintenance tasks. 

\noindent \textbf{LLM-based test repair and robust locators generation.} Our study shows promising results in using LLMs like ChatGPT for Web test repair (Finding 3).
Our manual analysis shows that it can generate correct repairs for the broken statements. Currently, the generated fixes mainly modifies and generate assertions, showing the promises of LLM-based test repair. In future, it is worthwhile to study using LLM for improving test repair technique that fixes broken assertions~\cite{daniel2011reassert}~\cite{yaraghi2024automated}. Another worthwhile future work is to use LLMs like ChatGPT to improve the locator robustness and incorporated into prior techniques like \emph{Robula+} ~\cite{robula2}. For robust locators generation, ChatGPT may be instructed to select different locators, prioritizing locators that are less fragile. 


\noindent \textbf{Improving reliability of ChatGPT output.} Prior study has expressed concerns regarding the tendency of ChatGPT to ``hallucinate'' when solving specific tasks~\cite{Feldt2023aa}, our workflow (that checks whether the explanations provided by ChatGPT are consistent along with the selected elements) 
shows promising results for improving the matching accuracy for certain combinations (Finding 7). 
Although our study are limited to Web UI test maintenance, we believe that our proposed way of checking for explanation consistencies is general and can be applied to improve the reliability of ChatGPT for other software maintenance tasks (e.g., test generation). 

\section{threats to validity}

\noindent \textbf{External Threats.} During the ground truth construction and evaluating our way of calculating EC, we mitigate potential bias by asking two annotators to manually construct and cross-validate the ``ground truth target element and patch''. The two annotators met to resolve any disagreement during the annotation, and further discussed until a consensus was reached. To reduce bias in selection of Web applications, we evaluate on a widely used dataset. As with prior evaluations, our findings may not generalize beyond the selected applications and tests in the dataset. 
 To encourage future research in Web UI element matching and repair, we also release our dataset. Due to limited resources and budget, we use the cost-effective GPT 3.5-Turbo model which may be less effective than newer models in Web test repair. Nevertheless, the extended input of a ChatGPT with a larger token limit (e.g., ChatGPT-16k) may still be insufficient for representing UI elements because for a Webpage, there are an average of 224 elements to be matched where the attribute information for each element occupies 101 tokens, 224*101=22624 tokens. 
 As ChatGPT performance may vary across settings, our experiments may not generalize beyond the studied settings, and fixing other forms of UI tests (we focus on Java Selenium Web UI tests~\cite{shariff2019improving}).
We mitigate this threat by reusing settings and suggestions in prior work (e.g., referring to OpenAI documentation for prompt design and using temperature 0.8 as in prior work ~\cite{fan2023automated}), and evaluating several test repair tools that use different algorithms (e.g., text-based and visual-based). As our study only evaluates the test repair capability of ChatGPT, the findings may not apply for other LLMs. Nevertheless, our suggested workflow of using LLMs to perform subsequent matching and rematching based on inconsistent explanations are still generally applicable. 

\noindent\textbf{Internal Threats.} Our experimental scripts may have bugs that can affect our results. To mitigate this threat, we made our scripts and results publicly available \cite{reproducibility}.

\noindent\textbf{Conclusion Threats.} Conclusion threats of our study include (1) overfitting of our dataset and (2) subjectivity of ground truth construction. To ensure that the matching and repair tasks do not overlap with the training dataset of ChatGPT to minimize the possibility of overfitting, we manually analyze the updated UI tests, and we have manually labeled and created a ground truth dataset that can be used to support future research in Web UI test repair. We mitigate the subjectivity of ground truth construction by cross-validating between two annotators during ground truth construction.  


\section{Conclusions}
This paper presents the first feasibility study that evaluates the effectiveness of
using prior Web UI repair techniques for initial matching and then using ChatGPT to perform subsequent matching to mitigate the bias in prioritization of attributes of prior approaches. To reduce hallucination in ChatGPT, we introduce an explanation validator that checks the consistency of the provided explanation, and gives hints to ChatGPT via a self-correction prompt to further improve its results. Our evaluation on a widely used dataset shows that the combinations with ChatGPT improve the effectiveness of prior techniques. Our study also reveals several findings and implications. As an initial study that focuses on LLM-based Web test repair, we hope that our study could shed light in improving future Web UI test repair approaches.

\section{Acknowledgements}
We greatly appreciate the effort of
Yuanzhang Lin for his help in annotating the ground truth fixes.
This work is supported by the Natural Sciences and Engineering Research Council of Canada (NSERC) Discovery Grants (RGPIN-2024-04301). 
\section*{Data Availability} 
To ensure reproducibility and advance future research in test repair, we made the dataset, scripts and experimental data publicly available on Github under an MIT license \cite{reproducibility}.



   \bibliographystyle{IEEEtran}
 \bibliography{sigproc}

\end{document}